\documentclass[useAMS,usenatbib,usegraphicx]{mn2e}
\title[Non variability of GRB080319 intervening absorbers]{Non variability of intervening absorbers observed in the
UVES spectra of the ``naked-eye'' GRB080319 \thanks{Based on
observations collected at the European Southern Observatory, ESO, the
VLT/Kueyen telescope, Paranal, Chile, in the framework of programs
080.A-0398}}
\author[V. D'Elia et al.]{V. D'Elia$^{1,2}$\thanks{E-mail:
delia@mporzio.astro.it}, F. Fiore$^1$, P. Goldoni$^{3,4}$, V. D'Odorico$^{5}$, S. Campana$^6$,
S. Covino$^6$, \newauthor P. D'Avanzo$^6$, E.J.A. Meurs$^{7,8}$, L. Norci$^8$, G. Tagliaferri$^6$ 
\\
$^1$ INAF-Osservatorio Astronomico di Roma, Via Frascati 33, I-00040 Monteporzio Catone, Italy;\\ 
$^2$ ASI-Science Data Centre, Via Galileo Galilei, I-00044 Frascati, Italy;\\ 
$^3$ Laboratoire Astroparticule et Cosmologie, 10 rue A. Domon et L. Duquet, 75205 Paris Cedex 13, France;\\
$^4$ Service d'Astrophysique, DSM/DAPNIA/SAp, CEA-Saclay, 91191 Gif-sur-Yvette, France;\\
$^5$ INAF, Osservatorio Astronomico di Trieste, Via Tiepolo 11, 34143 Trieste, Italy; \\
$^6$ INAF, Osservatorio Astronomico di Brera, via E. Bianchi 46, 23807 Merate (LC), Italy;\\ 
$^7$ School of Cosmic Physics, DIAS, 31 Fitzwilliam Place, Dublin 4, Ireland;\\
$^8$ School of Physical Sciences and NCPST, DCU, Glasnevin, Dublin 9, Ireland;\\
}
\begin{document}

\date{Accepted... Received...; in original form...}

\pagerange{\pageref{firstpage}--\pageref{lastpage}} \pubyear{2002}

\maketitle

\label{firstpage}

\begin{abstract}
The aim of this paper is to investigate the properties of the
intervening absorbers lying along the line of sight of Gamma-Ray Burst
(GRB) 080319B through the analysis of its optical absorption features.
To this purpose, we analyze a multi-epoch, high resolution
spectroscopic observations (R=40000, corresponding to $7.5$
km\,s$^{-1}$) of the optical afterglow of GRB080319B ($z=0.937$),
taken with UVES at the VLT. Thanks to the rapid response mode (RRM),
we observed the afterglow just 8m:30s after the GRB onset when the
magnitude was R $\sim 12$. This allowed us to obtain the best
signal-to-noise, high resolution spectrum of a GRB afterglow ever (S/N
per resolution element $\sim50$). Two further RRM and target of
opportunity observations were obtained starting $ 1.0$ and $2.4 $
hours after the event, respectively. Four {{Mg}{II}} absorption
systems lying along the line of sight to the afterglow have been
detected in the redshift range $0.5 < z < 0.8$, most of them showing a
complex structure featuring several components. Absorptions due to
{{Fe}{II}}, {{Mg}{I}} and {{Mn}{II}} are also present;
they appear in four, two and one intervening absorbers, respectively.
One out of four systems show a {{Mg}{II} $\lambda$2796} rest frame
equivalent width larger than 1\AA. This confirms the excess of strong
{{Mg}{II}} absorbers compared to quasars, with $dn/dz = 0.9$,
$\sim 4$ times larger than the one observed along quasar lines of
sight. In addition, the analysis of multi-epoch, high-resolution
spectra allowed us to exclude a significant variability in the column
density of the single components of each absorber.  Combining this
result with estimates of the size of the emitting region, we can
reject the hypothesis that the difference between GRB and QSO
{{Mg}{II}} absorbers is due to a different size of the emitting
regions.
\end{abstract}


\section{Introduction}

Intervening metal absorption lines have routinely been observed along
the line of sight to quasars (QSOs). Among these features, strong
{{Mg}{II} $\lambda$2796, $\lambda$2803} absorption doublets (with
equivalent width $EW>0.3$ \AA) represent good indicators of
metal-enriched gas associated with foreground galaxies. This is
because Magnesium is an $\alpha$ element produced by red giants and
dispersed in the interstellar medium (ISM) by supernovae and stellar
winds (Bergeron \& Boisse 1991, Steidel et al. 1994).  In particular,
such a strong doublet on a QSO sightline witnesses the presence of
$0.1-5 L_\star$ galaxies within 60 kpc projected distance and with
different morphological types (Churchill et al. 2005, Kacprzak et
al. 2008). In addition, the $EW>1$ \AA$\;$ {Mg}{II} intervening
absorbers are hosted within dark matter halos with characteristic
masses of $10^{11}-10^{12} M\odot$ (Bouche' et al. 2006) and $\sim
80\%$ are associated with Damped Lyman-alpha systems (DLAs, Rao et
al. 2006).

For a few hours after their onset, Gamma-Ray Burst (GRB) afterglows
are the brightest beacons in the far Universe: therefore they provide
an alternative and complementary way to QSOs to fully explore the
properties of high-z galaxies in which they are hosted (see Savaglio
2006; Prochaska et al. 2007) and of those along their sightlines. In
principle, the sightlines of QSOs and GRBs are expected to be
equivalent.  Prochter et al. (2006) found $\sim 7000$ strong ($EW>1
\AA\;$) {Mg}{II} intervening absorbers in $\sim50000$ QSO spectra
from the SDSS DR4, which corresponds to a redshift number density of
$dn/dz=0.24$.  Surprisingly, the same authors report the
identification of 14 {Mg}{II} systems along 14 GRB sightlines,
which translates into a redshift number density of
$dn/dz=0.90^{+0.83}_{-0.50}$ (99\% confidence interval), almost 4
times higher than the QSO one. On the other hand, the intervening
{{C}{IV}} absorbers for the two classes of sources do not show any
statistical difference (Sudilovsky et al. 2007, Tejos et al. 2007).

The reason for this discrepancy is still uncertain, and several
scenarios have been widely discussed (see e.g. Porciani et al. 2007
and Cucchiara et al. 2009). One of these interpretations requires a
different size of the source, namely, that QSO emitting regions are
larger than GRBs which in turn are comparable to the typical size of
the MgII clouds (Frank et al. 2007). In this scenario variability in
the column densities of the {{Mg}{II}} absorber is expected, since
the Lorentz factor decreases because the fireball decelerates on the
interstellar medium, and thus the GRB emission regions increase. A
{{Mg}{II}} variability in multi-epoch spectroscopy data on
GRB060206 was first claimed by the analysis by Hao et al. (2007), but
then disproved by Aoki et al. (2008) and Th\"one et al. (2008).

Occasionally, extremely bright optical transient emission is
associated with the GRB event, offering the superb opportunity to take
spectra of the afterglows with high resolution instruments. In a
fraction of these cases, multi-epoch, high resolution spectroscopy
with an adequate signal to noise ratio (S/N) can be obtained, allowing
to make detailed studies of the host galaxy ISM and to put strong
constraints on its physical parameters (see Vreeswijk et al. 2007 for
GRB060418, and D'Elia et al. 2009a for GRB080319B). Here we take
advantage of the high quality data collected for GRB083019B to make a
systematic study of this GRB sightline and to search for variability
of the {{Mg}{II}} intervening absorbers.
 
The paper is organized as follows. Section $2$  presents a short summary
of the GRB080319B detection and observations; Section $3$ describes the datasets
and data reduction; Section $4$ presents the full analysis of the
intervening systems identified in the spectra; finally in Section $5$
the results are discussed and conclusions are drawn.

\section{GRB080319B}

GRB080319B was discovered by the Burst Alert Telescope (BAT)
instrument on board Swift on 2008, March 19, at 06:12:49 UT. Swift
slewed to the target in less than 1 minute and a bright afterglow was
found by both the X-Ray Telescope (XRT) and UV-Optical Telescope
(UVOT) at RA = $14$h $31$m $40.7$s, Dec = +36$^o$ $18'$ $14.7"$
(Racusin et al. 2008a) with observations starting 60.5 and 175 s after
the trigger, respectively.

The field of GRB080319B was imaged by the "Pi of the Sky" apparatus
located at Las Campanas Observatory before, during and after the GRB
event (Cwiok et al. 2008). The field was also targetted by the robotic
telescope REM just 43 s after the BAT trigger (Covino et al. 2008a,
b). The TORTORA wide-field optical camera (12 cm diameter,
20$\times$25 deg FOV, TV-CCD, unfiltered) mounted on REM also imaged
the field before, during and after the GRB event with good temporal
resolution (Karpov et al. 2008).  These observations show that the GRB
reached the magnitudes $V = 5.3 $ about $20$ s and $H = 4.2$ about
$50$ s after the trigger.  This makes GRB080319B the brightest GRB
ever recorded at optical wavelengths (Racusin et al. 2008b, Bloom et
al. 2009).

\begin{table*}
\caption{\bf GRB080319B journal of observations and setups used}
{\footnotesize
\smallskip
\label{obs_log}
\begin{tabular}{|lccccccccc|}
\hline
\hline
Obs  & UT observation & T. from burst (s)& Setup (nm) & Wavelength (\AA) & Slit       & Resolution & Exp. (s) & S/N         & R mag \\
\hline
RRM 1  & 2008 Mar 19, 06:21:26 & 517   & Dic 2, 437   & 3760 - 4980      & 1''        & 40000      & 600      & $ \sim 30 $ & $12 - 13$\\
RRM 1  & 2008 Mar 19, 06:21:26 & 517   & Dic 2, 860   & 6700 - 10430     & 1''        & 40000      & 600      & $ \sim 50 $ & $12 - 13$\\
RRM 2  & 2008 Mar 19, 07:18:47 & 3441  & Dic 1, 346   & 3040 - 3870      & 1''        & 40000      & 1800     & $ \sim 7  $ & $16 - 17$\\
RRM 2  & 2008 Mar 19, 07:18:47 & 3441  & Dic 1, 580   & 4780 - 6810      & 1''        & 40000      & 1800     & $ \sim 12 $ & $16 - 17$\\
RRM 2  & 2008 Mar 19, 08:06:42 & 6833  & Dic 2, 437   & 3760 - 4980      & 1''        & 40000      & 1800     & $ \sim 7  $ & $16 - 17$\\
RRM 2  & 2008 Mar 19, 08:06:42 & 6833  & Dic 2, 860   & 6700 - 10430     & 1''        & 40000      & 1800     & $ \sim 12 $ & $16 - 17$\\
ToO    & 2008 Mar 19, 08:43:52 & 8546  & Dic 1, 346   & 3040 - 3870      & 1''        & 40000      & 1200     & $ \sim 5  $ & $16 - 17$\\
ToO    & 2008 Mar 19, 08:43:52 & 8546  & Dic 1, 580   & 4780 - 6810      & 1''        & 40000      & 1200     & $ \sim 8  $ & $16 - 17$\\
ToO    & 2008 Mar 19, 09:07:18 & 10478 & Dic 2, 437   & 3760 - 4980      & 1''        & 40000      & 1200     & $ \sim 5  $ & $16 - 17$\\
ToO    & 2008 Mar 19, 09:07:18 & 10478 & Dic 2, 860   & 6700 - 10430     & 1''        & 40000      & 1200     & $ \sim 8  $ & $16 - 17$\\

\hline
\end{tabular}
}
\end{table*}

\section{Observations and data reduction}

We observed the bright afterglow of GRB080319B in the framework of the
ESO program 080.A-0398 with the VLT/UVES (Dekker et al. 2000).  The
Observation Log and the setups used are reported in Table 1. Both
UVES dichroics, as well as the red and the blue arms, were used.

The first, 10min observation, was performed in Rapid Response Mode
(RRM) and started just 8m:30s after the GRB event, when the afterglow
was extremely bright ($R_{Mag}=12-13$). This provided a S/N=$30 - 50$ per
resolution element. Two more UVES observations followed, the first one
was again in RRM mode, activated in the framework of program 080.D-0526
and starting 1.0 hours after the GRB event. The second one was a Target
of Opportunity (ToO), starting 2.4 hours after the GRB, see Table 1.  A
slit width of 1'' has been used in all the observations; this
corresponds to a spectral resolution of $R = 40000$, or $7.5$
km\,s$^{-1}$.

Data reduction was carried out by using the UVES pipeline (Ballester
et al. 2000). The final useful spectra extend from $\sim 3800$~\AA{}
to $\sim 9500$~\AA.  The spectra were normalized to the continuum,
which was evaluated by fitting the data with cubic splines, after the
removal of the absorption features. Finally, the noise spectrum, used
to determine the errors on the best fit line parameters, was
calculated from the real, background-subtracted spectra using
line-free regions. This takes into account both statistical and
systematic errors in the pipeline processing and background
subtraction.

\section{The GRB080319B sightline}

An analysis of the GRB080319B UVES spectra reveals at least 5
absorbing systems along the GRB line of sight.  The presence of
excited {{Fe}{II}} and {{Ni}{II}} features in the higher
redshift system is a clear indication that its gas belongs or is close
to the host galaxy of the GRB. We will not discuss the host galaxy
absorber, since a detailed study has been reported by D'Elia et
al. (2009a); but instead we will now concentrate on the intervening
systems.

The four intervening absorbers identified have redshifts in the range
$0.5 - 0.8$. Each system features {{Mg}{II}} absorption.
{{Fe}{II}}, {{Mg}{I}} and {{Mn}{II}} are also present;
they appear in four, two and one intervening absorbers,
respectively. The analysis of these systems is often intricate due to
the complexity of the absorption lines of the spectrum, which in
several cases cannot easily be fit with a single line profile.  
The presence of several components is indicative of clumpy gas in the
intervening absorbers, similarly to what observed for the circumburst
environment of many GRBs (See e.g., D'Elia et al. 2007, Piranomonte et
al. 2009, D'Elia et al. 2009a,b) and for the {{Mg}{II}}
intervening absorbers along the QSO sightlines (Churchill \& Vogt
2001). Whenever a system can not be fit by a single line profile, it
can be interpreted both as a clumpy system with a complex velocity
structure or as many systems lying at different redshifts. This is why
we say that GRB080319B has at least four intervening systems. We
arbitrarily chose to consider as a single system two or more
absorption features closer than $500$ km s$^{-1}$. Tab. 2 summarizes
the characteristics of the intervening absorbers. Column 2 gives the
heliocentric redshift, column 3 the absorbing elements and ions,
column 4 the total width of the system, column 5 the rest frame
equivalent width (EW$_{rf}$) of the {{Mg}{II}$\lambda$2796} line,
column 6 the number of components necessary to adequately fit each
system. The intervening absorbers are ordered with decreasing
redshift. For multiple component systems the $z$ reference values have
been arbitrarily placed to be coincident with the component indicated
in column 7; for single component absorbers the redshift is defined by
the best fit value of the central absorption line wavelength.

\begin{table*}
\caption{\bf Absorption systems along the GRB080319B sightline}
{\footnotesize
\smallskip
\label{obs_log}
\begin{tabular}{|l|cccccc|}
\hline
\hline
System & redshift         & species   & Width (km s$^{-1}$) & {{Mg}{II} $\lambda$2796} EW$_{rf}^*$ (\AA)         & \# of components & z reference component     \\
\hline
1      & $0.76046$        & {Mg}{II}, {Fe}{II}                            & $90$              &$0.121\pm0.007$                         & $4$    &    2nd    \\
2      & $0.71468$        & {Mg}{II}, {Mg}{I}, {Fe}{II}               & $400$             &$1.448\pm0.007$                         & $10$   &    7th    \\
3      & $0.56578$        & {Mg}{II}, {Fe}{II}                            & $30$              &$0.083\pm0.006$                         & $1$    &    1st    \\
4      & $0.53035$        & {Mg}{II}, {Mg}{I}, {Fe}{II}, {Mn}{II} & $100$             &$0.62\pm0.01$                           & $6$    &    3rd    \\

\hline
\end{tabular}

$^*$ Rest frame values, errors are at $1\sigma$ confidence level.
}
\end{table*}

The next subsections report the analysis for each intervening
absorber.  The fitting procedure for the absorption systems has been
carried out as follows. First of all, we analyzed the first epoch
spectrum, that with the highest signal-to-noise ratio. The spectrum
was analyzed in the MIDAS environment using the {\sc fitlyman}
procedure (Fontana \& Ballester 1995). 
The line profile fitting is usually performed using a Voigt function.  
Each Voigt profile has basically three free parameters: the central
wavelength of the transition, the column density of the absorbing
species and the doppler parameter $b$ of the gas. A single component
treatment is often inadequate, reflecting the complexity of the
intervening systems, as noted before. We thus fit the data several
times with different numbers of components, in order to minimize the
reduced $\chi ^2$ values. {Mg}{II}, which appears in all systems,
is the one with a larger velocity spread, so it was used to guide the
identification of all components.  The other species, when present,
allowed us to best constrain the central wavelength positions in the
regions where the {Mg}{II} results to be strongly saturated. The
central wavelengths and the $b$ parameters have been kept fixed among
the different species, unless otherwise stated. Once a satisfactory
fit to the first epoch spectrum was obtained, we turned to the
analysis of the other two epochs. Again, the central wavelengths and
the $b$ parameters were fixed to the first epoch results. To increase
the S/N of the later epoch observations, we added them (the second and third
spectra) and repeated the fits to the coadded spectrum. Tables 3 to 6
report the results of our analysis for the intervening absorbers 1 to
4, respectively.  In particular, column 2 shows the number of features
contributing to the fit for each species, column 3 reports the epoch
to which the data refers, and the following ones the column densities
of each component for that specific element or ion. Components are
identified with progressive numbers for decreasing redshifts (or
decreasing wavelengths, i.e., the higher the wavelength or the
positive velocity shift, the lower the component number). Errors are
the formal $1 \sigma$ uncertainties given by {\sc fitlyman}; upper
limits have the $90\% $ level confidence.

\subsection{The intervening system 1}

This is the system with the lowest number of absorption features. In
fact, only the {{Mg}{II} $\lambda$2796, $\lambda$2803} doublet and
the {{Fe}{II} $\lambda$2382} line are present at
$z=0.76046$. Despite this, the structure of the system, which spans a
velocity range of $\sim 90$ km s$^{-1}$ is quite complex, and a four
component model is necessary in order to obtain a reasonable fit to
the data. Fig. 1 shows the absorption due to the {Mg}{II} doublet
and the {{Fe}{II} $\lambda$2382} line, together with the results
of the four component fit. The {{Fe}{II} $\lambda$2382} line is
only present in the fourth component of the first observation.  The S/N
ratio of the following observations just allows to set upper limits for
this feature. Table 3 reports the column densities for each
component. We can rule out a variability of both the {Mg}{II} and
{Fe}{II} lines, since all the column densities are consistent
within the $2 \sigma$ level.

The non variability of the {Mg}{II} features is also shown in
Fig. 2, which displays the {{Mg}{II} $\lambda$2796} features for
the three epochs.

\begin{figure}
\centering
\includegraphics[angle=-90,width=9cm]{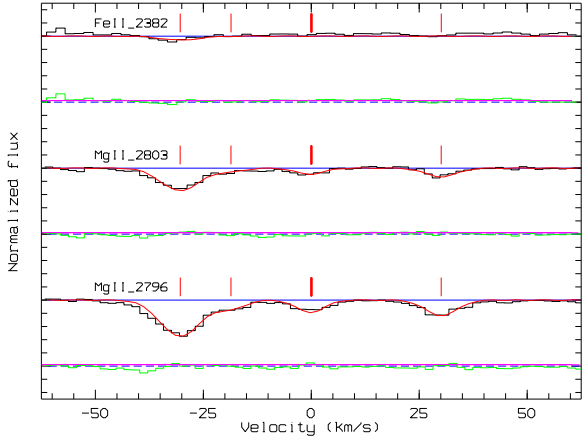}
\caption{The {{Mg}{II} $\lambda$2796, $\lambda$2803} absorption
doublet and the {{Fe}{II} $\lambda$2382} line of the first
intervening system, for the first epoch spectrum. Solid lines
represent our four Voigt components, best fit model. Vertical lines
identify the velocity of each component with respect to the zero point
arbitrarily placed at $z=0.76046$ and coincident with component II.}
\end{figure}

\begin{table*}
\caption{\bf Column densities for the absorption system 1}
{\footnotesize
\smallskip
\begin{tabular}{|lc|ccccc|}
\hline
\hline
Species        & Trans.          & Obs.& I (+30 km s$^{-1}$) & II (0 km s$^{-1}$) & III (-20 km s$^{-1}$) & IV (-30 km s$^{-1}$)\\
\hline						
Mg II         &  $\lambda$2796  & 1   & $11.83 \pm 0.02$   & $11.91 \pm 0.16$     & $10.56 \pm 0.07   $ & $12.37 \pm 0.01   $\\
              &  $\lambda$2803  & 2   & $11.75 \pm 0.09$   & $ < 11.7       $     & $  <11.7          $ & $12.33 \pm 0.04   $\\
              &                 & 3   & $ < 12.1       $   & $ < 12.1       $     & $  <12.1          $ & $12.22 \pm 0.07   $\\
              &                 &2+3  & $11.72 \pm 0.08$   & $ < 11.7       $     & $  <11.7          $ & $12.29 \pm 0.03   $\\
\hline
Fe II         &  $\lambda$2382  & 1   & $ < 11.5       $   & $ < 11.5       $     & $  <11.5          $ & $11.53 \pm 0.09   $\\
              &                 & 2   & $ < 12.7       $   & $ < 12.7       $     & $  <12.7          $ & $ < 12.7          $\\
              &                 & 3   & $ < 13.1       $   & $ < 13.1       $     & $  <13.1          $ & $ < 13.1          $\\
              &                 &2+3  & $ < 12.7       $   & $ < 12.7       $     & $  <12.7          $ & $ < 12.7          $\\
\hline
\end{tabular}

All values are logarithmic cm$^{-2}$
}
\end{table*}

\begin{figure}
\centering
\begin{tabular}{cc}
\includegraphics[width=6cm, angle=-90]{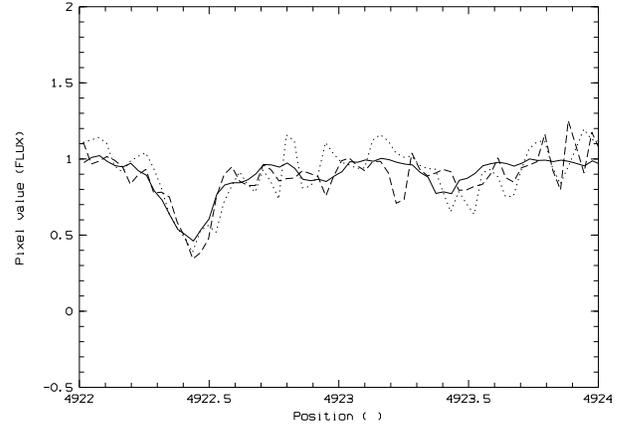}
\end{tabular}
\caption{A comparison between the {{Mg}{II} $\lambda$2796} optical
depth in the three UVES observation epochs, for the first intervening
system. Solid line refers to the first epoch spectrum (8m30s after the
Swift trigger), dashed line to the second epoch spectrum (1.0 hours
after the GRB event), and dotted line to the third epoch spectrum (2.4
hours after the GRB event).}
\end{figure}

\subsection{The intervening system 2}

This is the most complex system, which spans the largest velocity
range ($\sim 400$ km s$^{-1}$). A lot of absorption features are
present at the redshift of $0.71468$, namely: the {{Mg}{II}
$\lambda$2796, $\lambda$2803} doublet, {Mg}{I} ($\lambda$2852) and
{Fe}{II} in several flavours ($\lambda$2344, $\lambda$2374,
$\lambda$2382, $\lambda$2586 and $\lambda$2600). The first epoch
spectrum required a ten component fit in order to obtain a good
modeling to the data. Fig. 3 shows the absorption due to the
{Mg}{II} doublet and {{Mg}{I} $\lambda$2852}, while Fig. 4
displays all the features of {Fe}{II}; in both figures the resuls
of the ten component fit are also plotted. Table 4 reports the column
densities for each component. No variability is detected in the three
epochs within the $3 \sigma $ uncertainty. This rules out column
density variability for the intervening system 2. The only exception
is represented by the second component of the {Mg}{II}. Anyway,
this specific component is strongly saturated, so the values reported
in the corresponding column of Table 4 may not be entirely reliable.

The non variability of the features belonging to the second system can
also be seen in Figs. 5-7, which display the {{Mg}{II} $\lambda$2796},
{{Mg}{I} $\lambda$2852} and {{Fe}{II} $\lambda$2600} features for the
three epochs, respectively. This system is the only one showing an
EW$_{MgII,rf}>1$ so we also computed the EWs of the absorption
features for comparison with other works.  The {{Mg}{II}$\lambda$2796}
EW does not vary between the three observations at the $2\sigma$
confidence level, and a variability of this feature, if present, is
less than 10\% at the $3\sigma$ confidence level. The {{Fe}{II}
  $\lambda$2600} and {{Mg}{I} $\lambda$2852} rest frame EWs are
EW$_{FeII,rf}=0.708\pm0.007$ and EW$_{MgI,rf}=0.289\pm0.007$,
respectively ($1\sigma$ confidence). These values are not different
from that observed along other GRB sightlines (see e.g.  Cucchiara et
al. 2009).

\begin{figure}
\centering
\includegraphics[angle=-90,width=9cm]{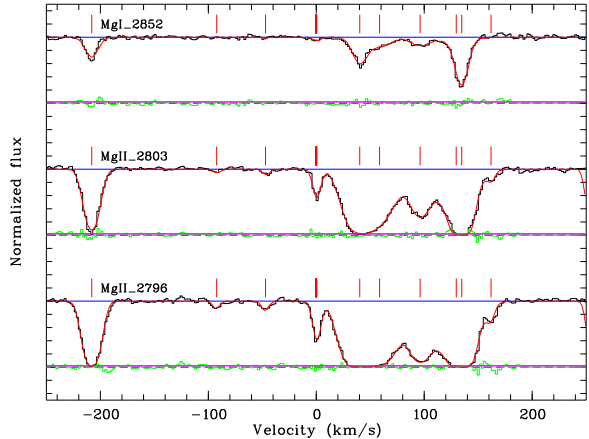}
\caption{The {{Mg}{II} $\lambda$2796, $\lambda$2803} doublet and
the {{Mg}{I} $\lambda$2852} feature of the second intervening
system, for the first epoch spectrum. Solid lines represent our ten
Voigt components, best fit model. Vertical lines identify the velocity
of each component with respect to the zero point arbitrarily placed at
$z=0.71468$ and coincident with component VII.}
\end{figure}

\begin{figure}
\centering
\includegraphics[angle=-90,width=9cm]{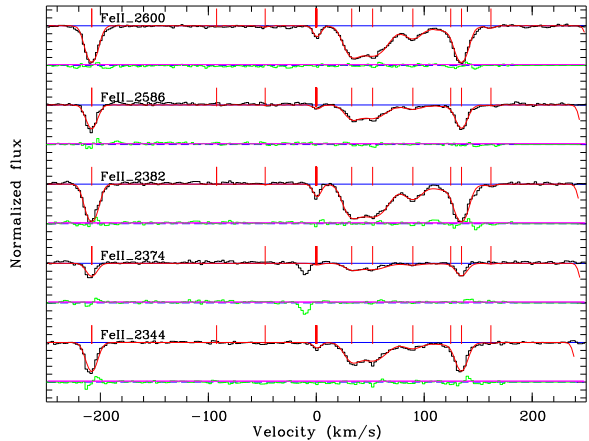}
\caption{All the {Fe}{II} features of the second intervening
system, for the first epoch spectrum. Solid lines represent our ten
Voigt components, best fit model. Vertical lines identify the velocity
of each component with respect to the zero point arbitrarily placed at
$z=0.71468$ and coincident with component VII.}
\end{figure}

\begin{table*}
\caption{\bf Column densities for the absorption system 2}
{\footnotesize
\smallskip
\begin{tabular}{|lc|cccccc|}
\hline
\hline
Species        & Trans.          & Obs.& I (160 km s$^{-1}$)& II (135 km s$^{-1}$)& III (125 km s$^{-1}$)& IV (90 km s$^{-1}$)& V (50 km s$^{-1}$)  \\
\hline			      				
              &                 &     & VI (30 km s$^{-1}$)& VII (0 km s$^{-1}$)&VIII (-50 km s$^{-1}$)& IX (-100 km s$^{-1}$)& X (-210 km s$^{-1}$)\\
\hline			      				
Mg II         &  $\lambda$2796  & 1   & $13.84 \pm 0.05$   & $15.45 \pm 0.08$    & $13.05 \pm  0.02  $  & $13.04 \pm 0.02   $ & $13.29 \pm 0.02   $ \\
              &  $\lambda$2803  & 2   & $13.79 \pm 0.12$   & $16.01 \pm 0.06$    & $13.03 \pm  0.04  $  & $13.07 \pm 0.13   $ & $13.32 \pm 0.04   $ \\
              &                 & 3   & $ < 12.2       $   & $14.83 \pm 0.12$    & $13.07 \pm  0.04  $  & $13.21 \pm 0.19   $ & $13.19 \pm 0.11   $ \\
              &                 &2+3  & $13.57 \pm 0.16$   & $15.11 \pm 0.30$    & $13.24 \pm  0.06  $  & $12.99 \pm 0.04   $ & $13.19 \pm 0.08   $ \\
\hline			      				
              &                 & 1   & $13.65 \pm 0.02$   & $12.57 \pm 0.03$    & $12.66 \pm  0.49  $  & $12.56 \pm 0.34   $ & $13.44 \pm 0.03   $ \\
              &                 & 2   & $13.64 \pm 0.12$   & $12.50 \pm 0.06$    & $12.37 \pm  0.28  $  & $13.03 \pm 1.95   $ & $13.57 \pm 0.05   $ \\
              &                 & 3   & $13.56 \pm 0.10$   & $12.37 \pm 0.09$    & $ < 12.1          $  & $12.27 \pm 0.64   $ & $13.53 \pm 0.04   $ \\
              &                 &2+3  & $13.70 \pm 0.07$   & $12.43 \pm 0.06$    & $ < 11.9          $  & $12.78 \pm 0.23   $ & $13.54 \pm 0.05   $ \\
\hline			      				
Mg I          &  $\lambda$2852  & 1   & $ < 10.5       $   & $12.18 \pm 0.01$    & $11.19 \pm  0.07  $  & $11.29 \pm 0.03   $ & $11.45 \pm 0.03   $ \\
              &                 & 2   & $ < 11.2       $   & $12.10 \pm 0.04$    & $ < 11.2          $  & $ < 11.2          $ & $11.29 \pm 0.11   $ \\
              &                 & 3   & $ < 11.6       $   & $12.10 \pm 0.03$    & $ < 11.6          $  & $ < 11.6          $ & $ < 11.6          $ \\
              &                 &2+3  & $ < 11.2       $   & $12.08 \pm 0.05$    & $11.34 \pm  0.14  $  & $ < 11.2          $ & $11.13 \pm 0.12   $ \\
\hline			      				
              &                 & 1   & $11.77 \pm 0.01$   & $10.63 \pm 0.09$    & $ < 10.5          $  & $ < 10.5          $ & $11.45 \pm 0.03   $ \\
              &                 & 2   & $11.77 \pm 0.04$   & $ < 11.2       $    & $ < 11.2          $  & $ < 11.2          $ & $11.49 \pm 0.06   $ \\
              &                 & 3   & $11.75 \pm 0.03$   & $ < 11.6       $    & $ < 11.6          $  & $ < 11.6          $ & $ < 11.6          $ \\
              &                 &2+3  & $11.78 \pm 0.05$   & $ < 11.2       $    & $ < 11.2          $  & $ < 11.2          $ & $11.58 \pm 0.05   $ \\
\hline
FeII&$\lambda$2344, $\lambda$2374& 1  & $13.33 \pm 0.08$   & $13.43 \pm 0.01$    & $12.40 \pm  0.03  $  & $12.68 \pm 0.01   $ & $13.40 \pm 0.02   $ \\
              &  $\lambda$2382  & 2   & $13.79 \pm 0.18$   & $13.35 \pm 0.04$    & $12.52 \pm  0.09  $  & $12.64 \pm 0.04   $ & $13.43 \pm 0.02   $ \\
              &  $\lambda$2586  & 3   & $13.44 \pm 0.56$   & $13.39 \pm 0.06$    & $ < 12.7          $  & $12.80 \pm 0.05   $ & $13.42 \pm 0.03   $ \\
              &  $\lambda$2600  &2+3  & $13.77 \pm 0.17$   & $13.35 \pm 0.03$    & $12.53 \pm  0.08  $  & $12.71 \pm 0.03   $ & $13.41 \pm 0.02   $ \\
\hline
              &                 & 1   & $13.15 \pm 0.01$   & $12.35 \pm 0.02$    & $ < 11.7          $  & $ < 11.7          $ & $13.48 \pm 0.01   $ \\
              &                 & 2   & $13.09 \pm 0.04$   & $ < 12.4       $    & $ < 12.4          $  & $ < 12.4          $ & $13.40 \pm 0.04   $ \\
              &                 & 3   & $13.12 \pm 0.05$   & $ < 12.7       $    & $ < 12.7          $  & $ < 12.7          $ & $13.43 \pm 0.05   $ \\
              &                 &2+3  & $13.12 \pm 0.08$   & $12.18 \pm 0.08$    & $ < 12.1          $  & $ < 12.1          $ & $13.41 \pm 0.03   $ \\
\hline

\end{tabular}

All values are logarithmic cm$^{-2}$
}
\end{table*}

\begin{figure}
\centering
\begin{tabular}{cc}
\includegraphics[width=6cm, angle=-90]{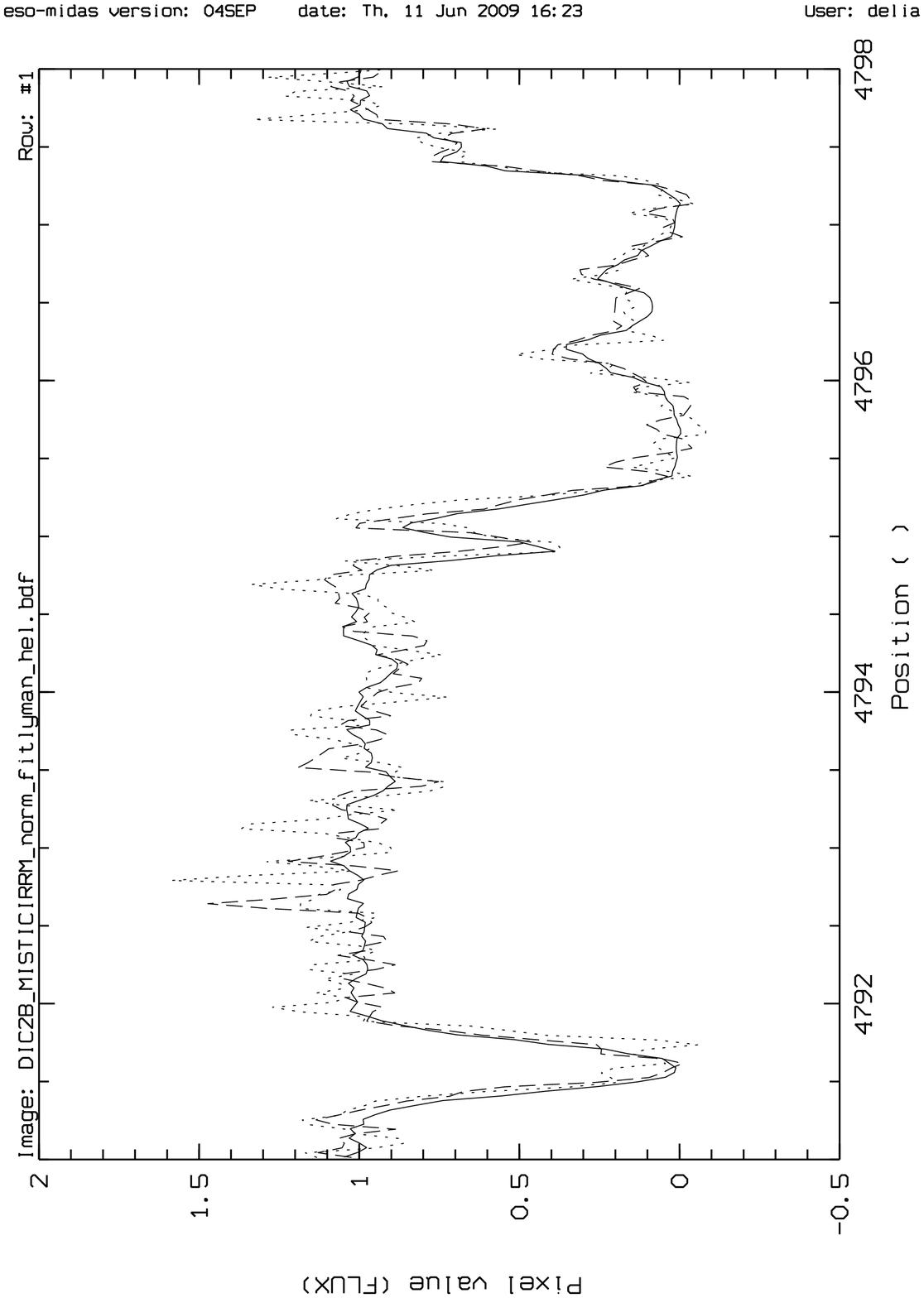}
\end{tabular}
\caption{A comparison between the {{Mg}{II} $\lambda$2796} optical
depth in the three UVES observation epochs, for the second intervening
system. Solid line refers to the first epoch spectrum (8m30s after the
Swift trigger), dashed line to the second epoch spectrum (1.0 hours
after the GRB event), and dotted line to the third epoch spectrum (2.4
hours after the GRB event).}
\end{figure}

\begin{figure}
\centering
\begin{tabular}{cc}
\includegraphics[width=6cm, angle=-90]{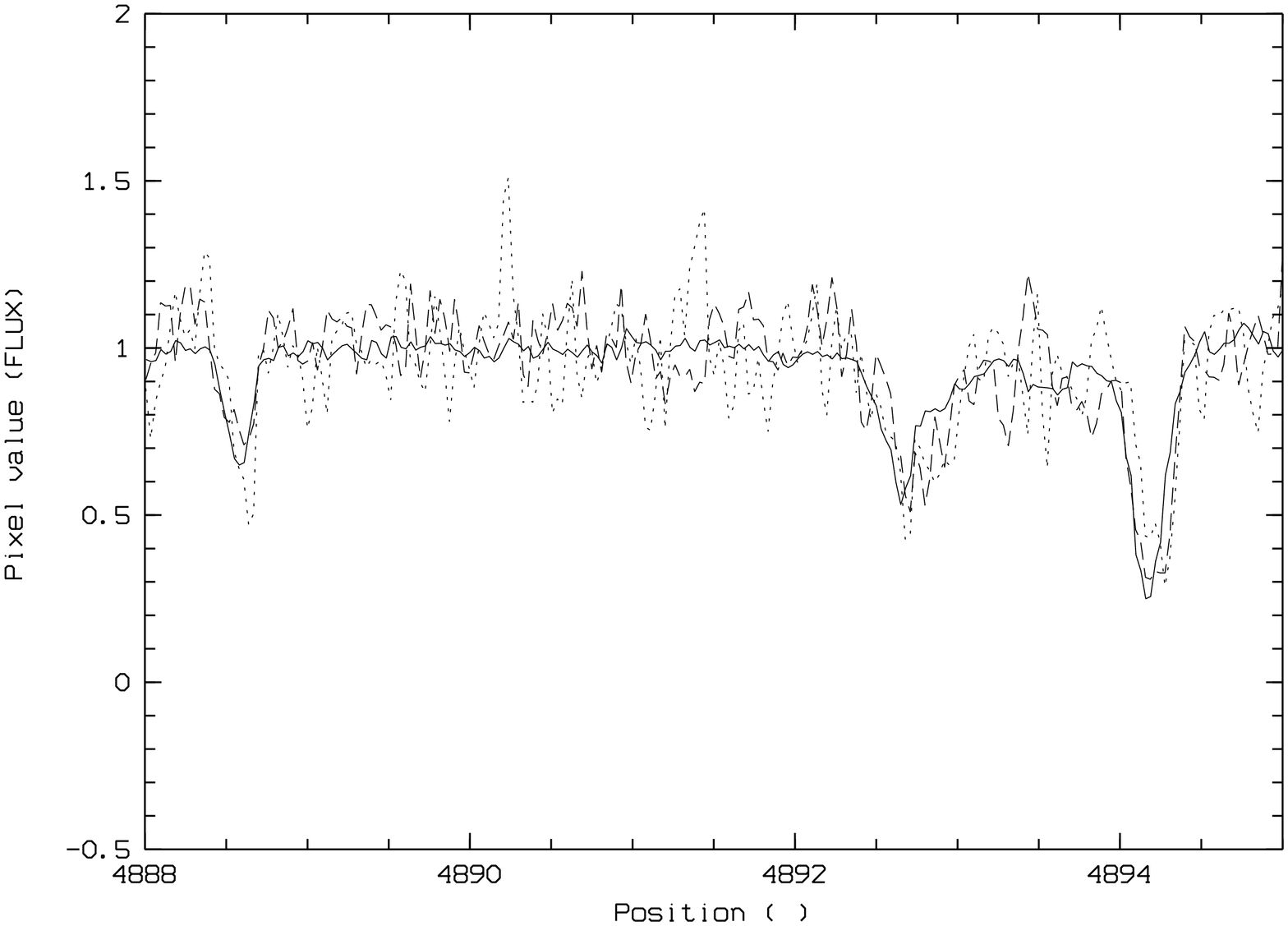}
\end{tabular}
\caption{A comparison between the {{Mg}{I} $\lambda$2852} optical
depth in the three UVES observation epochs, for the second intervening
system. Solid line refers to the first epoch spectrum (8m30s after the
Swift trigger), dashed line to the second epoch spectrum (1.0 hours
after the GRB event), and dotted line to the third epoch spectrum (2.4
hours after the GRB event).}
\end{figure}

\begin{figure}
\centering
\begin{tabular}{cc}
\includegraphics[width=6cm, angle=-90]{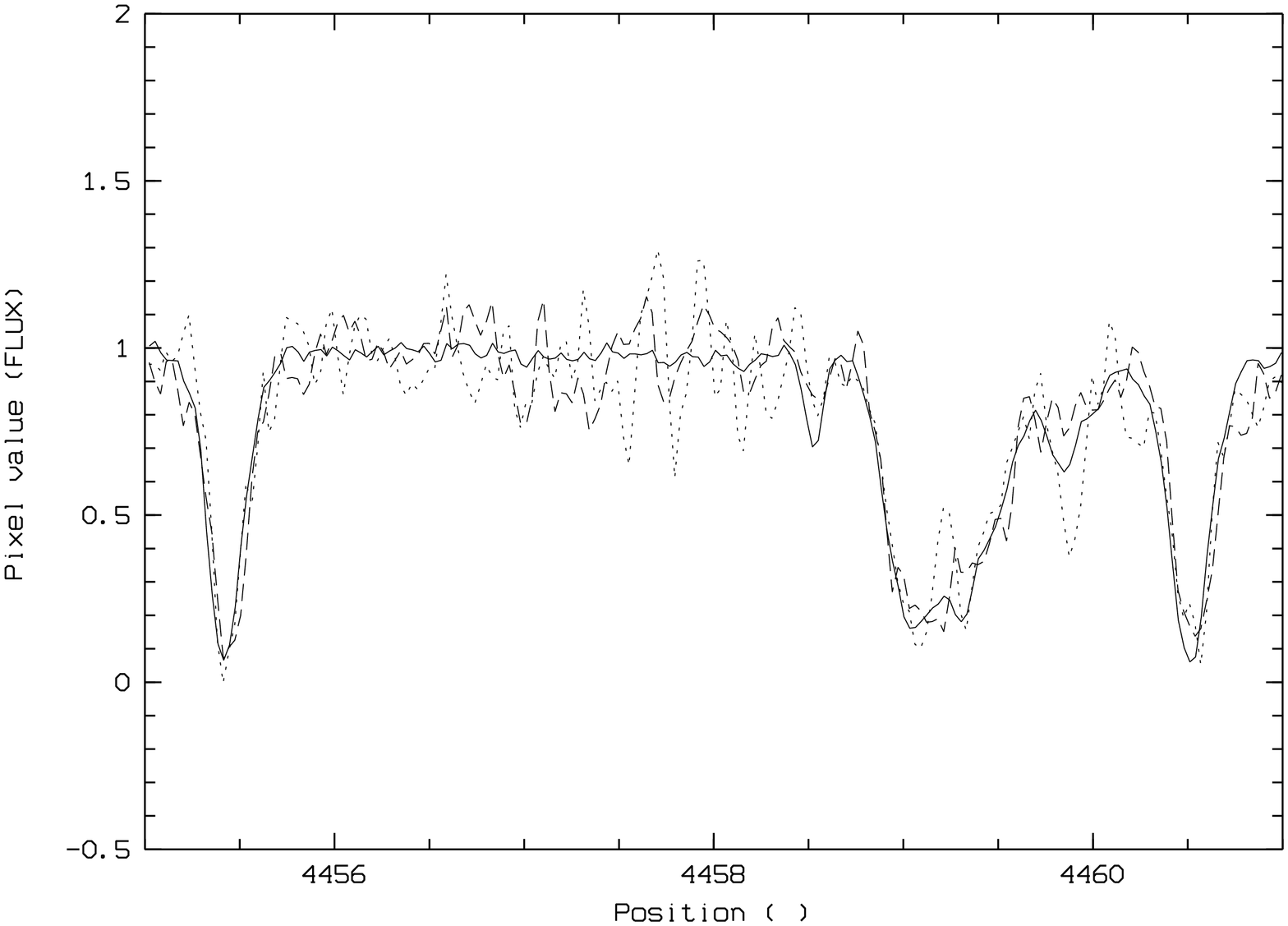}
\end{tabular}
\caption{A comparison between the {{Fe}{II} $\lambda$2600} optical
depth in the three UVES observation epochs, for the second intervening
system. Solid line refers to the first epoch spectrum (8m30s after the
Swift trigger), dashed line to the second epoch spectrum (1.0 hours
after the GRB event), and dotted line to the third epoch spectrum (2.4
hours after the GRB event).}
\end{figure}

\subsection{The intervening system 3}

This is the simplest system among the intervening absorbers. Its
velocity dispersion is just $\sim 30$ km s$^{-1}$ and only two ions
appear in the spectrum at the redshift of $z=0.56578$: {Mg}{II},
with the classical $\lambda$2796, $\lambda$2803 doublet and
{Fe}{II} ($\lambda$2586 and $\lambda$2600). All the three epoch
spectra are well fit by a single component Voigt profile. Fig. 8 shows
the absorption from the {Mg}{II} and {Fe}{II} features,
together with our fit to the data. Table 5 reports the column
densities measured for the three epochs.  We can rule out a
variability of both {Mg}{II} and {Fe}{II} lines, since the
column densities of the three epochs are consistent within the $3
\sigma$ level. 

The non variability of the features belonging to the third system is 
also shown in Figs 9 and 10, which display the {{Mg}{II}
$\lambda$2796} and {{Fe}{II} $\lambda$2600} features for the three
epochs, respectively.


\begin{figure}
\centering
\includegraphics[angle=-90,width=9cm]{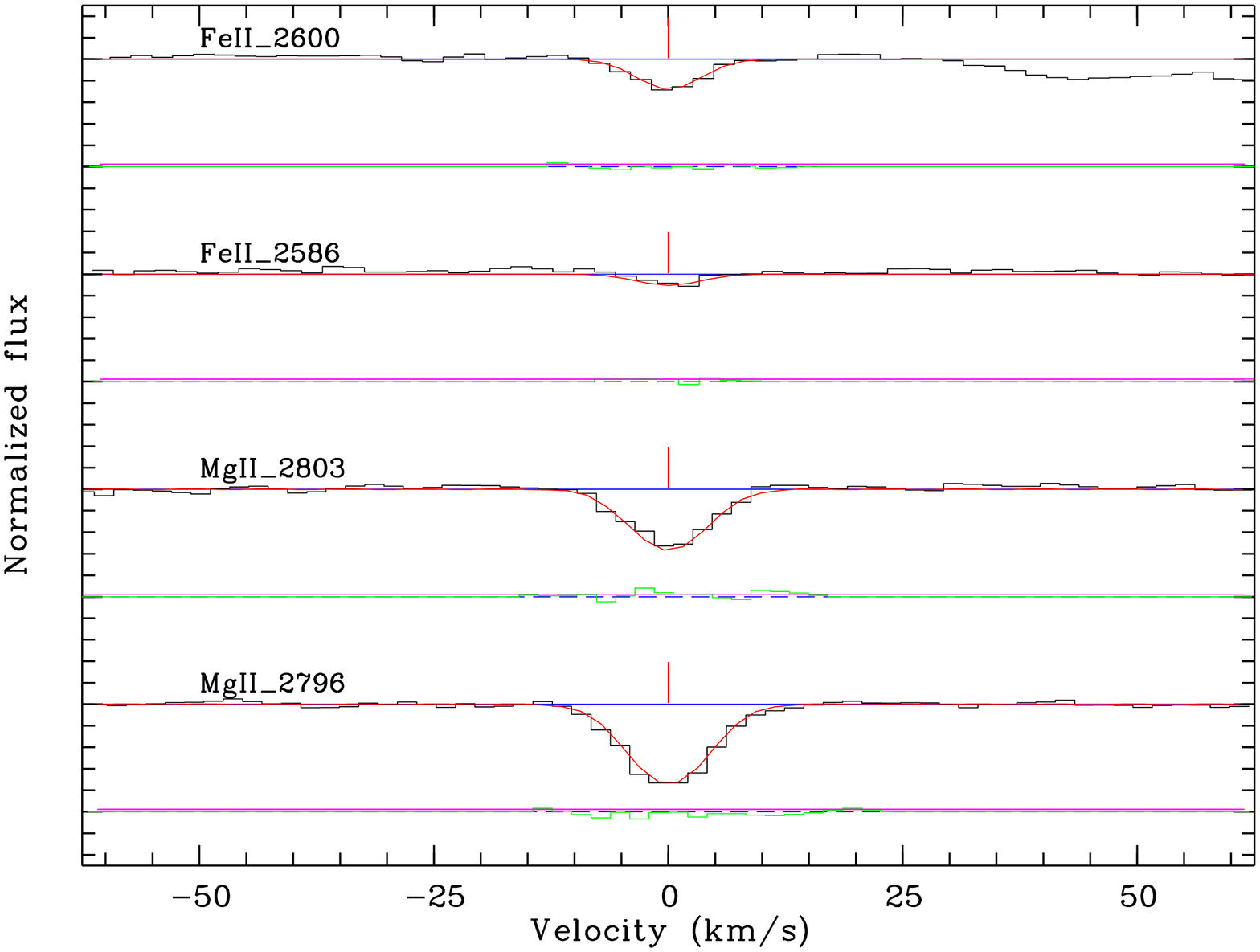}
\caption{The {{Mg}{II} $\lambda$2796, $\lambda$2803} doublet and
the {{Fe}{II} $\lambda$2586, $\lambda$2600} absorption features of
the third intervening system, for the first epoch spectrum. Solid
lines represent our Voigt profile, best fit model. The vertical
line identifies the position of the central wavelength of the voigtian.
}
\end{figure}

\begin{table*}
\caption{\bf Column densities for the absorption system 3}
{\footnotesize
\smallskip
\begin{tabular}{|lc|cc|}
\hline
\hline
Species        & Trans.          & Obs.& I (0 km s$^{-1}$)  \\
\hline
Mg II         &  $\lambda$2796  & 1   & $12.73 \pm 0.02$    \\
              &  $\lambda$2803  & 2   & $12.57 \pm 0.05$    \\
              &                 & 3   & $12.71 \pm 0.09$    \\
              &                 &2+3  & $12.63 \pm 0.05$    \\
\hline
Fe II         &  $\lambda$2586  & 1   & $12.40 \pm 0.04$    \\
              &  $\lambda$2600  & 2   & $12.28 \pm 0.18$    \\
              &                 & 3   & $ < 12.7       $    \\
              &                 &2+3  & $12.18 \pm 0.15$    \\
\hline
\end{tabular}

All values are logarithmic cm$^{-2}$
}
\end{table*}

\begin{figure}
\centering
\begin{tabular}{cc}
\includegraphics[width=6cm, angle=-90]{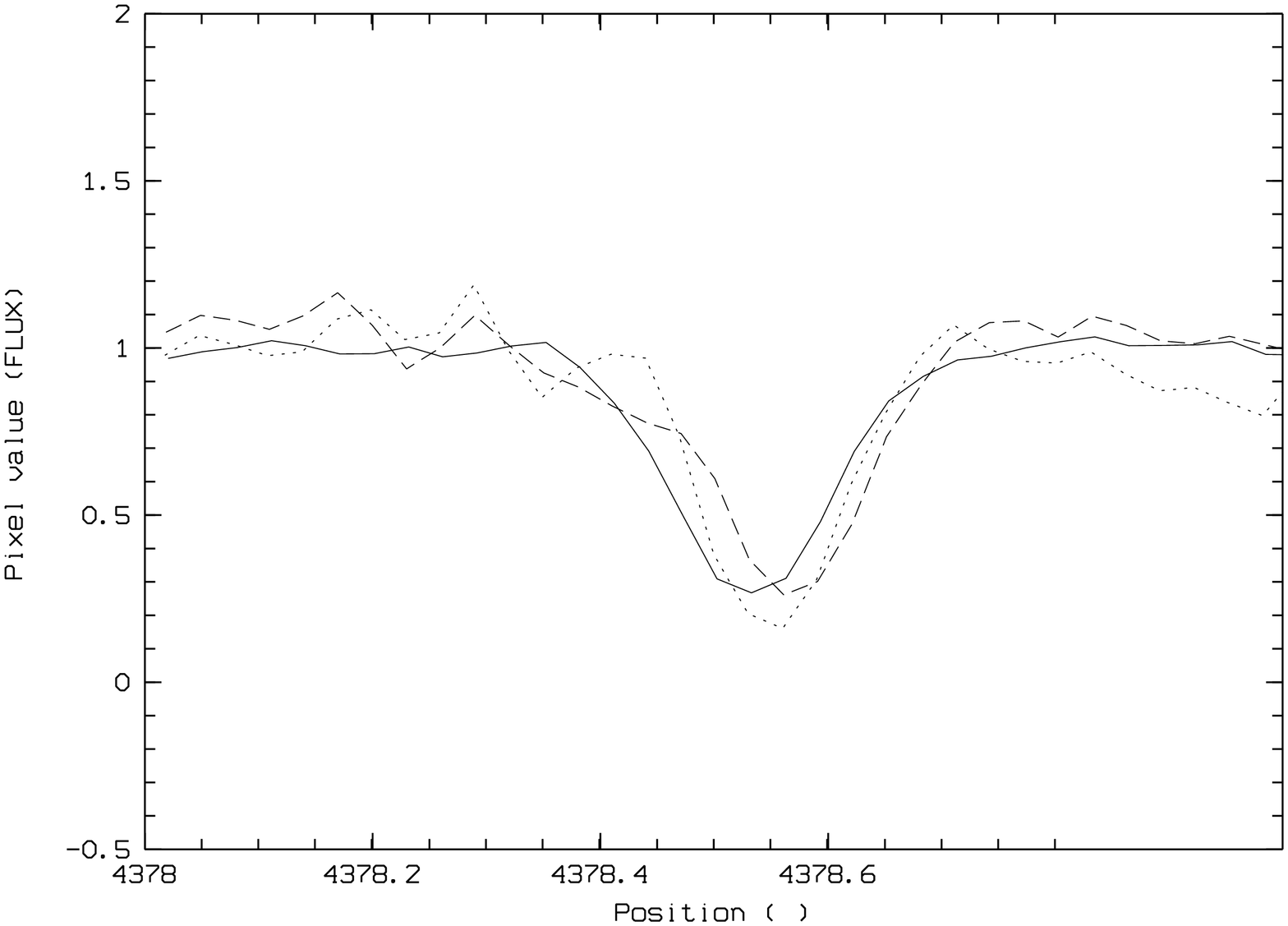}
\end{tabular}
\caption{A comparison between the {{Mg}{II} $\lambda$2796} optical
depth in the three UVES observation epochs, for the third intervening
system. Solid line refers to the first epoch spectrum (8m30s after the
Swift trigger), dashed line to the second epoch spectrum (1.0 hours
after the GRB event), and dotted line to the third epoch spectrum (2.4
hours after the GRB event). The slight wavelength shift of the second
and third epoch spectra with respect to the first one ($\sim 0.04$\AA$\;$
or $<3$ km/s, lower than the spectral resolution, see also fig. 10) is
possibly an offset that comes out from the data reduction and not a
real effect.}
\end{figure}

\begin{figure}
\centering
\begin{tabular}{cc}
\includegraphics[width=6cm, angle=-90]{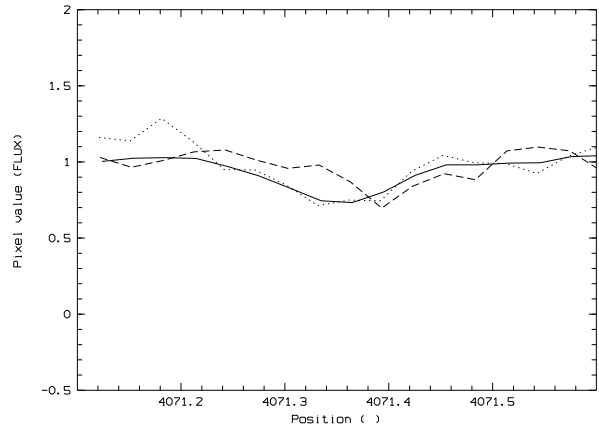}
\end{tabular}
\caption{A comparison between the {{Fe}{II} $\lambda$2600} optical
depth in the three UVES observation epochs, for the third intervening
system. Solid line refers to the first epoch spectrum (8m30s after the
Swift trigger), dashed line to the second epoch spectrum (1.0 hours
after the GRB event), and dotted line to the third epoch spectrum (2.4
hours after the GRB event).}
\end{figure}

\subsection{The intervening system 4}

This is the system which shows the greatest number of species. Three
ions, {Mg}{II} ($\lambda$2796, $\lambda$2803), {Fe}{II}
($\lambda$2586, $\lambda$2600), {Mn}{II} ($\lambda$2576,
$\lambda$2594) and the neutral {Mg}{I} ($\lambda$2852) appear in
the spectrum at the redshift of $z=0.53035$. The velocity dispersion
of the system is $\sim 100$ km s$^{-1}$. Since both {Mg}{II}
features are strongly saturated, we used {Mg}{I} and {Fe}{II}
to guide the identification of the components. The spectrum around
these features results to be well fit by a six component
model. Fig. 11 shows the absorptions from the {Mg}{II} and
{Mg}{I} features, while Fig. 12 those from {Fe}{II} and
{Mn}{II}. These figures also display our six component fit to the
data. Table 6 reports the column densities measured for the three
epochs divided by components. Variability can be excluded for
{Mg}{I}, {Fe}{II} and {Mn}{II} whose components are
consistent in the three epochs within the $3 \sigma$ uncertainty.
{Mg}{II} seems not to behave this way, but most of the components of
its features are strongly saturated, so the corresponding column
density values reported in the table are less reliable. Components 1
and 6 are not saturated (the former is not saturated only in the
{Mg}{II} $\lambda$2803 feature), and they are consistent with no
variability within the $3 \sigma$ uncertainty. It is worth noting,
however, that component 6 has a positive detection in the first
observation only, while in the other epochs just upper limits can be
set. Saturation is also present in component 2 and 3 of the
{Fe}{II}, but just in the first observation.

The non variability of the features belonging to the fouth system
is also shown in Figs. 13-15, which display the {{Mg}{II}
$\lambda$2796}, {{Mg}{I} $\lambda$2852} and {{Fe}{II}
$\lambda$2600}, features for the three epochs, respectively.

\begin{figure}
\centering
\includegraphics[angle=-90,width=9cm]{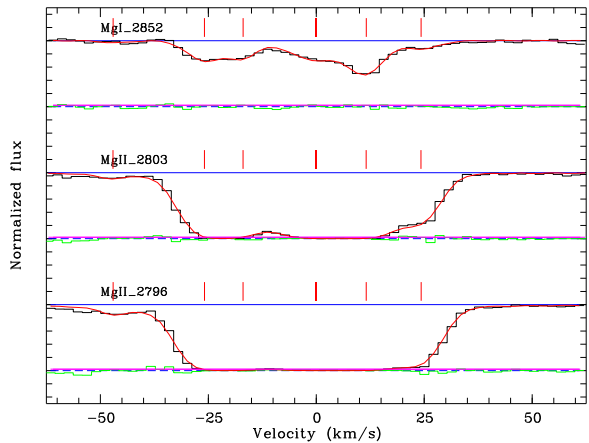}
\caption{The {{Mg}{II} $\lambda$2796, $\lambda$2803} doublet and
the {{Mg}{I} $\lambda$2852} absorption features of the fourth
intervening system, for the first epoch spectrum. Solid lines
represent our six component, best fit model.  Vertical lines identify
the velocity of each component with respect to the zero point
arbitrarily placed at $z=0.53035$ and coincident with component III.  }
\end{figure}

\begin{figure}
\centering
\includegraphics[angle=-90,width=9cm]{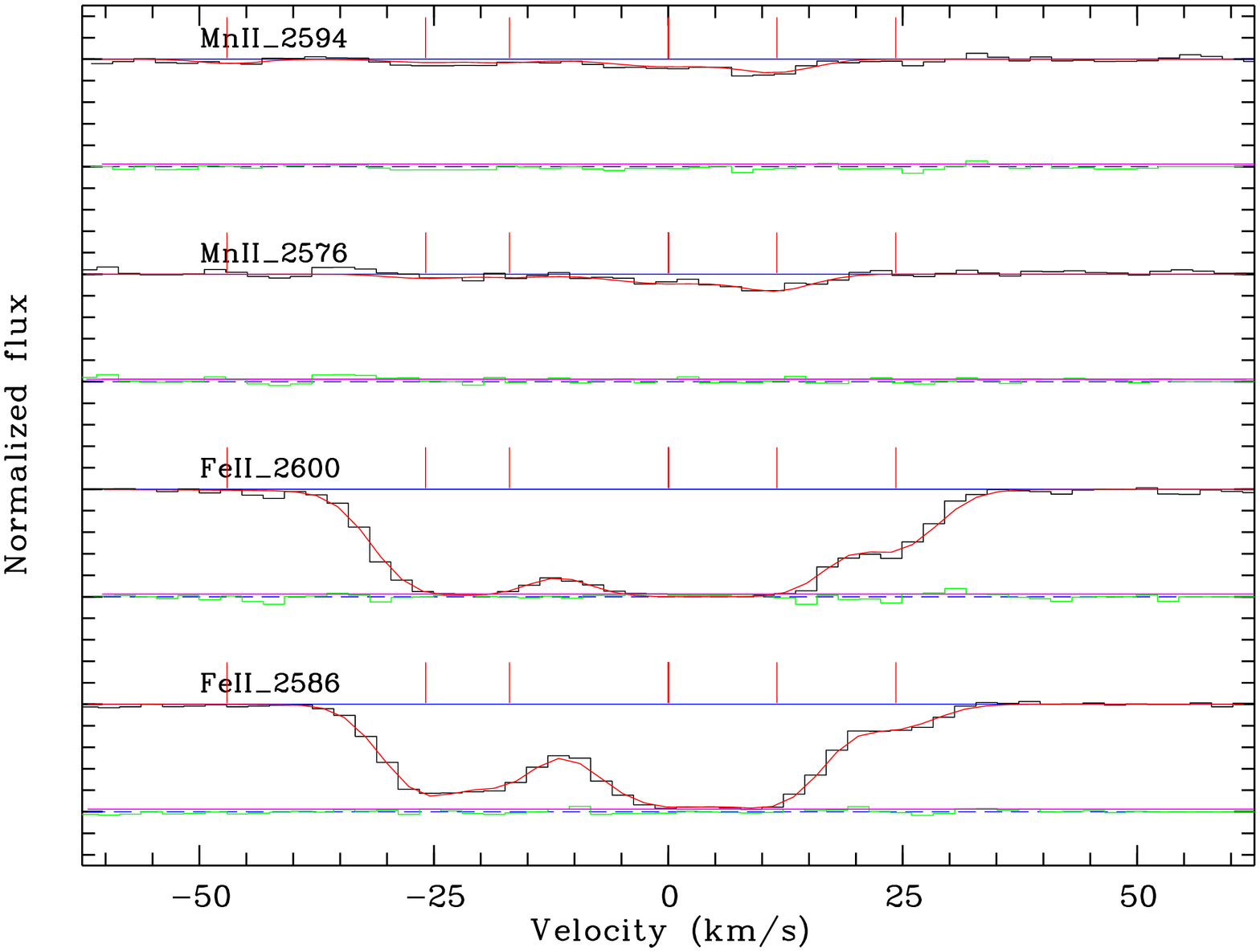}
\caption{The {{Fe}{II} $\lambda$2586, $\lambda$2600} and the
{{Mn}{II} $\lambda$2576, $\lambda$2594} absorption features of the
fourth intervening system, for the first epoch spectrum. Solid lines
represent our six component, best fit model.  Vertical lines identify
the velocity of each component with respect to the zero point
arbitrarily placed at $z=0.53035$ and coincident with component III.  }
\end{figure}

\begin{table*}
\caption{\bf Column densities for the absorption system 4}
{\footnotesize
\smallskip
\begin{tabular}{|lc|ccccccc|}
\hline
\hline
Species        & Trans.      & Obs.& I (25 km s$^{-1}$) & II (10 km s$^{-1}$) & III (0 km s$^{-1}$) & IV (-15 km s$^{-1}$) &  V (-25 km s$^{-1}$)  & VI (-50 km s$^{-1}$)   \\
\hline
Mg II         &$\lambda$2796& 1   & $12.94 \pm 0.02$   & $15.00S \pm 0.15$   & $15.25S \pm  0.04 $  & $13.22S \pm 1.71  $& $14.61S \pm 0.04  $ & $11.40 \pm 0.04   $  \\
              &$\lambda$2803& 2   & $12.75 \pm 0.09$   & $15.24S \pm 0.56$   & $16.13S \pm  0.15 $  & $15.68S \pm 0.22  $& $13.02S \pm 0.22  $ & $ < 11.9          $  \\
              &             & 3   & $12.79 \pm 0.10$   & $15.81S \pm 0.06$   & $14.75S \pm  0.13 $  & $14.94S \pm 0.12  $& $13.91S \pm 0.18  $ & $ < 12.1          $  \\
              &             &2+3  & $12.71 \pm 0.08$   & $15.93S \pm 0.07$   & $13.83S \pm  0.16 $  & $15.39S \pm 0.24  $& $13.91S \pm 0.17  $ & $ < 11.8          $  \\
\hline
Mg I          &$\lambda$2852& 1   & $11.17 \pm 0.06$   & $12.48 \pm 0.09$    & $11.67 \pm  0.03  $  & $11.44 \pm 0.03   $& $11.45 \pm 0.48   $ & $10.82 \pm 1.86   $  \\
              &             & 2   & $ < 11.2       $   & $11.96 \pm 0.30$    & $11.64 \pm  0.08  $  & $11.67 \pm 0.08   $& $ < 11.2          $ & $ < 11.2          $  \\
              &             & 3   & $ < 11.5       $   & $12.96 \pm 0.42$    & $11.79 \pm  0.09  $  & $11.53 \pm 0.13   $& $ < 11.5          $ & $ < 11.5          $  \\
              &             &2+3  & $ < 11.1       $   & $13.02 \pm 0.80$    & $11.71 \pm  0.07  $  & $11.59 \pm 0.07   $& $ < 11.1          $ & $ < 11.1          $  \\
\hline
Fe II         &$\lambda$2586& 1   & $12.80 \pm 0.06$   & $15.74S \pm 0.09$   & $14.38S \pm  0.09 $  & $13.65 \pm 0.07   $& $13.51 \pm 0.15   $ & $ < 11.8          $  \\
              &$\lambda$2600& 2   & $12.75 \pm 0.10$   & $13.67 \pm 0.47$    & $14.03 \pm  0.13  $  & $13.68 \pm 0.17   $& $13.75 \pm 0.50   $ & $ < 12.3          $  \\
              &             & 3   & $12.85 \pm 0.23$   & $15.06 \pm 0.81$    & $14.09 \pm  0.36  $  & $13.69 \pm 0.24   $& $13.33 \pm 0.24   $ & $ < 12.5          $  \\
              &             &2+3  & $12.81 \pm 0.12$   & $14.33 \pm 0.60$    & $14.05 \pm  0.16  $  & $13.67 \pm 0.19   $& $13.54 \pm 0.49   $ & $ < 12.2          $  \\
\hline
Mn II         &$\lambda$2576& 1   & $ < 11.4       $   & $11.80 \pm 0.03$    & $11.78 \pm  0.03  $  & $11.56 \pm 0.16   $& $ < 11.4          $ & $ < 11.4          $  \\
              &$\lambda$2594& 2   & $ < 12.4       $   & $ < 12.4       $    & $ < 12.4          $  & $ < 12.4          $& $ < 12.4          $ & $ < 12.4          $  \\
              &             & 3   & $ < 12.7       $   & $ < 12.7       $    & $ < 12.7          $  & $ < 12.7          $& $ < 12.7          $ & $ < 12.7          $  \\
              &             &2+3  & $ < 12.2       $   & $ < 12.2       $    & $ < 12.2          $  & $ < 12.2          $& $ < 12.2          $ & $ < 12.2          $  \\
\hline
\end{tabular}

All values are logarithmic cm$^{-2}$
}
\end{table*}

\begin{figure}
\centering
\begin{tabular}{cc}
\includegraphics[width=6cm, angle=-90]{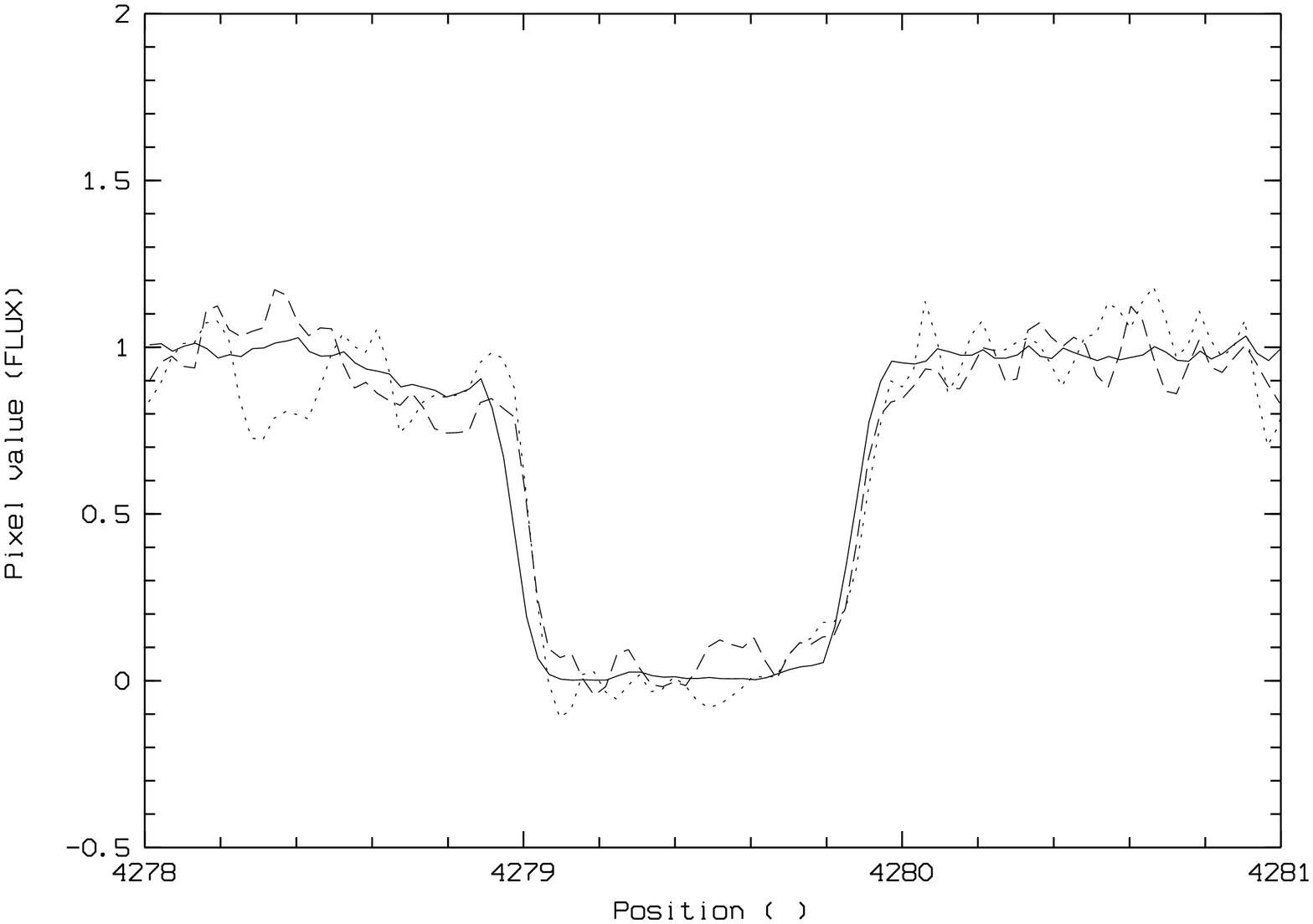}
\end{tabular}
\caption{A comparison between the {{Mg}{II} $\lambda$2796} optical
depth in the three UVES observation epochs, for the fourth intervening
system. Solid line refers to the first epoch spectrum (8m30s after the
Swift trigger), dashed line to the second epoch spectrum (1.0 hours
after the GRB event), and dotted line to the third epoch spectrum (2.4
hours after the GRB event). The $\sim 0.04$\AA$\;$ frequency shift (see
also figs. 14 and 15) is present also in this system, confirming that
its nature is not physical, but due to the reduction process.}
\end{figure}

\begin{figure}
\centering
\begin{tabular}{cc}
\includegraphics[width=6cm, angle=-90]{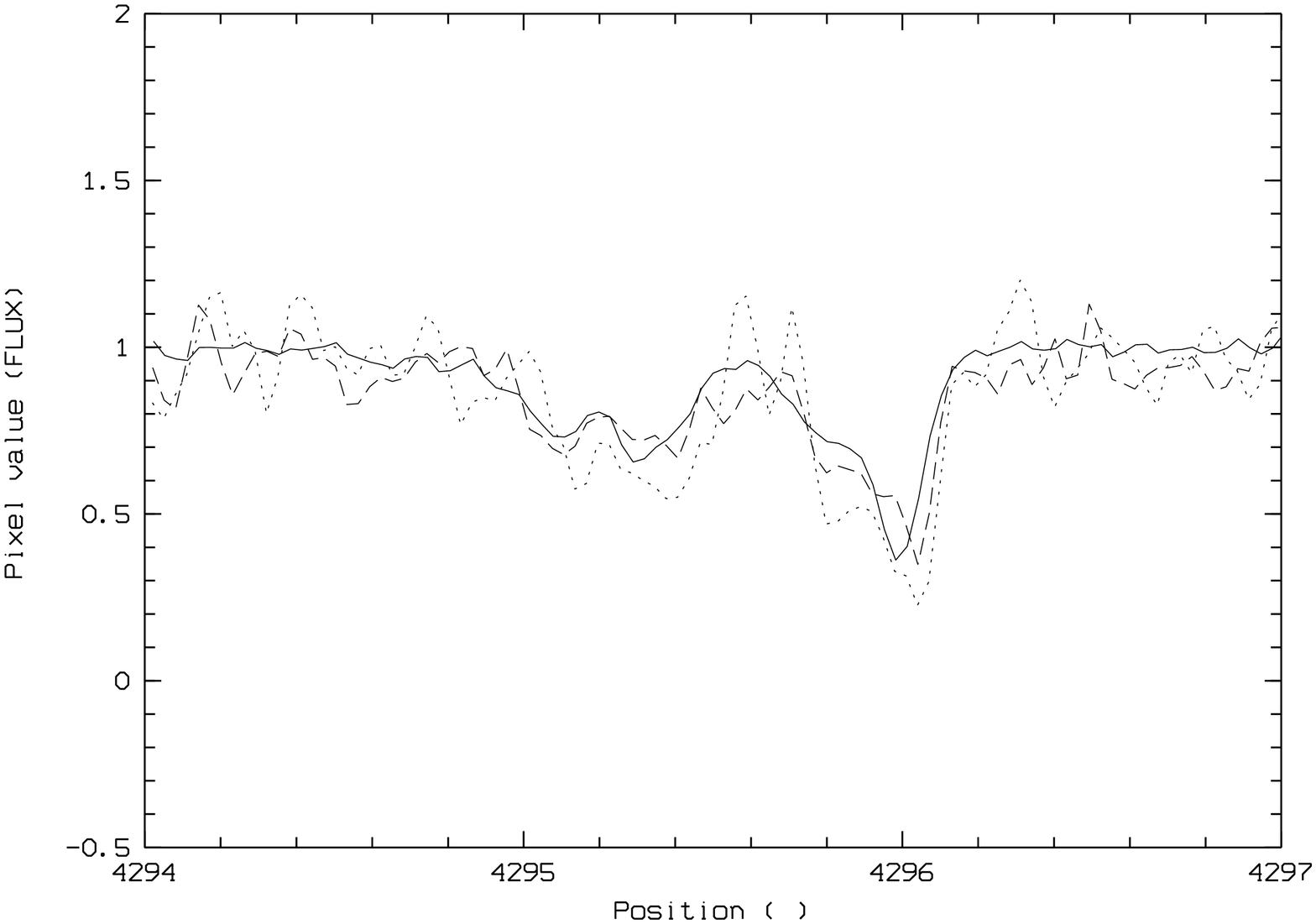}
\end{tabular}
\caption{A comparison between the {{Mg}{I} $\lambda$2852} optical
depth in the three UVES observation epochs, for the fourth intervening
system. Solid line refers to the first epoch spectrum (8m30s after the
Swift trigger), dashed line to the second epoch spectrum (1.0 hours
after the GRB event), and dotted line to the third epoch spectrum (2.4
hours after the GRB event).}
\end{figure}

\begin{figure}
\centering
\begin{tabular}{cc}
\includegraphics[width=6cm, angle=-90]{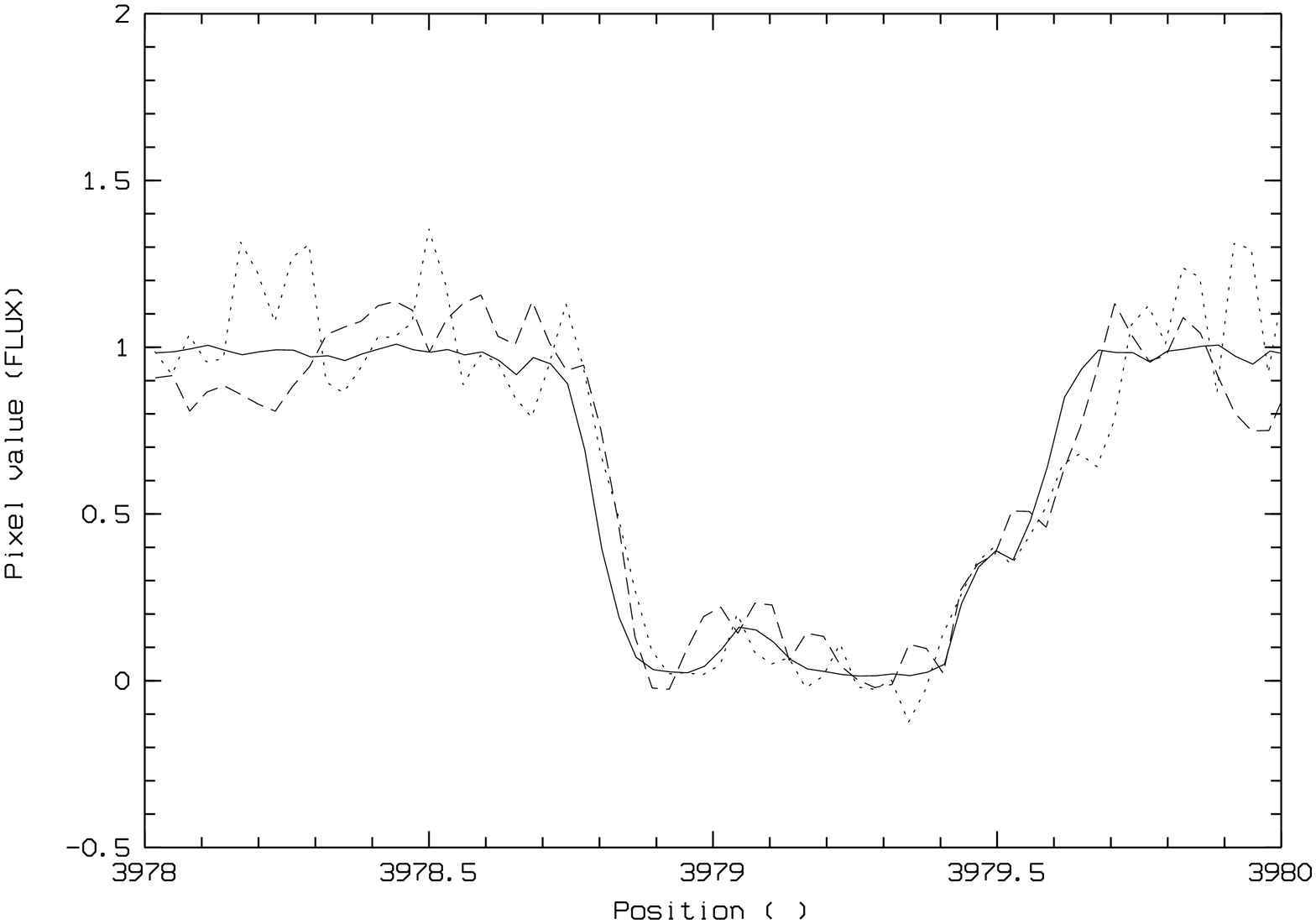}
\end{tabular}
\caption{A comparison between the {{Fe}{II} $\lambda$2600} optical
depth in the three UVES observation epochs, for the fourth intervening
system. Solid line refers to the first epoch spectrum (8m30s after the
Swift trigger), dashed line to the second epoch spectrum (1.0 hours
after the GRB event), and dotted line to the third epoch spectrum (2.4
hours after the GRB event).}
\end{figure}


\section{Conclusions}

In this paper we present high resolution (R=40000, corresponding to
7.5 km/s) spectroscopy of the optical afterglow of the ``naked-eye''
Gamma Ray Burst GRB080319B ($z=0.937$), observed by UVES at the VLT in
three different epochs, starting $\sim 8.30$ minutes, $\sim 1$ and
$\sim 2.4$ hours after the trigger, respectively.
                                                                        
We concentrate here on the intervening absorbers along the GRB
sightline, since the absorption features in the vicinity of the
afterglow have already been studied and presented in D'Elia et
al. (2009a). Our spectral coverage allows to analyze a redshift path
for the {{Mg}{II} $\lambda$2796} in the range $z=0.36 - 0.937$.
We are sensitive to lines with EW down to $0.015$ (at the
$2\sigma$ confidence level), thanks to the extremely high S/N of the
first RRM observation. At least four intervening systems between z =
0.8 and z = 0.5 have been identified. Most of them show a complex
structure constituted by several components, similar to that of the
intervening absorbers along the QSO sightlines (Churchill \& Vogt
2001). All systems feature the {{Mg}{II} $\lambda$2796,
$\lambda$2803} doublet. {Fe}{II}, {Mg}{I} and {Mn}{II}
lines are detected in four, two and one systems, respectively.

Prochter et al. (2006) claimed that the incidence of strong
{{Mg}{II}} absorbers along GRB sight lines is nearly four times
higher than that along the line of sight to QSOs. They analyzed the
spectra of 14 GRB optical afterglows, finding on average one
intervening system per afterglow with equivalent width $>1$ \AA. The
GRB080319B sightline confirms this trend, since the rest frame EW of
the {{Mg}{II} $\lambda$2796} absorption line is $>1$\AA$\;$ in one
system.
 
Structured intervening systems are expected if the discrepancy between
QSOs and GRBs intervening absorbers is due to a different size of the
source, namely, if QSO emitting regions are larger than GRBs which in
turn are comparable to the typical size of the{Mg}{II} clouds
(Frank et al. 2007). According to their estimation, the QSO beam size
must be larger than 3 $\times$ 10$^{16}$ cm while for the GRB beam
they suggest a size around a few $\times$ 10$^{15}$ cm. To observe
this effect, the MgII absorber should then be patchy with a clump size
around 10$^{16}$ cm. Nevertheless, in this scenario, variability in the
column densities of the {{Mg}{II}} absorbers along the GRB
sightlines is expected.  {{Mg}{II}} variability in multi-epoch
spectroscopy data on GRB060206 was first claimed by the analysis by
Hao et al. (2007), but then disproved by Aoki et al. (2008) and
Th\"one et al. (2008). This work confirms the lack of variability in
the column densities of the GRB intervening absorbers, in the specific
case of GRB080319B. In particular, the EW of the {{Mg}{II}
$\lambda$2796} absorption line is consistent within $2\sigma$ in the
three UVES observations. Given the S/N ratio of our data, we can
conclude that a variability of this feature, if present, is less than
10\% at the $3\sigma$ confidence level. This upper limit is three
times smaller than the ones that can be estimated for GRB060206 using
the data by Aoki et al. (2008). Racusin et al. (2008b) modeled the
radiation of GRB080319B as coming from a structured jet constituted by
an inner, narrow jet ($\theta_n = 0.2^o$) and an outer, wider jet
($\theta_w = 4^o$). They interpreted the optical emission at $T < T_0
+ 800$s ($T_0$ being the burst detection time) as produced by the
reverse shock of the inner jet, while that at $T > T_0 + 800$ as the
signature of the forward shock of the outer jet. Since the first UVES
observation starts at $T \sim T_0 + 500$ while the second and the
third ones start more than $1$hr later, a strong increase in the size
of the emitting region is expected from this scenario.  In more
detail, the dimension of the emitting region is $\sim 10^{17}$ cm at
the beginning of the GRB event and $\sim 10^{18}$ cm a few hours later
(Pandey et al. 2009, Racusin et al. 2008b and Kumar \& Panaitescu
2008). Although we cannot be more quantitative, a variation of a
factor of ten in the dimension of the emitting region would imply a
considerable reduction of the intervening column densities, but it is
not detected. Indeed, time resolved, high resolution spectroscopy
allows us for the first time to exclude a significant ($>3\sigma$)
variability even by considering the single components of each
intervening system alone.

Porciani et al. (2007) investigated several possible explanations for
the strong {{Mg}{II}} excess, namely, dust obscuration bias,
clustering of the absorbers, different beam sizes of the sources,
multiband magnification bias of GRBs, association of the absorbers
with the GRB event or with the circumburst environment. They concluded
that none of these effects alone can explain the observed difference,
but maybe the combination of two or more of them can reduce the
significance of the discrepancy. We can take advantage of the
dimensions of the emitting source in GRB080319B and the lack of
variability in the intervening absorbers to characterize the absorbers
along this GRB sightline. Since these dimensions are in the range
$10^{17} - 10^{18} $ cm, the {Mg}{II} (and other species) clouds
cannot have clumps or structures smaller than $0.1 - 1$ pc, otherwise
we should observe variability in their absorption. This is not
obvious, since Ding et al. (2003) show that smaller intervening
absorbers are present along the line of sight to QSOs.

Cucchiara et al. (2009) compared 81 QSOs and 6 GRB lines of sight
obtained with UVES. They found no significant evidence to support a
difference between the two absorber populations, concluding that a
possible explanation for the {{Mg}{II}} excess could be intrinsic
to the GRB environment. A similarity between the narrow lines of the
GRB intervening absorbers and that produced by the material ejected in
the accretion disk winds of QSOs has been pointed out. Nevertheless,
the high velocities required by the intervening redshifts ($v/c \sim
0.2 - 0.4$) and the lack of fine structure features in these systems
represent a strong weakness of this picture. In addition, Tejos et
al. (2009) recently found no evidence for a similar excess considering
the weak ($0.07 < EW < 1$\AA) {Mg}{II} systems along the line of
sight of 8 GRB observed with echelle spectrographs. These authors tend
to exclude an intrinsic nature of the discrepancy, since it would
result in an excess in the weak systems too. They suggest that the
best explanation available at present could be gravitational lensing
bias due to lensing by the host galaxy of the absorber. Indeed the
strong {Mg}{II} system along the GRB080319B sightline has a
velocity of $\sim 36300$ km/s with respect to the GRB. The data do not
show evidence either of broad profiles and/or partial coverage (see
e.g. D'Odorico et al. 2004), so this absorber is likely an intervening
system.

Combining the data by Prochter et al. (2006) with the results from the
analysis in this paper and that in D'Elia et al. (2009b) regarding the
line of sight to GRB080330, we obtain 16 strong (EW$>1$\AA)
{Mg}{II} intervening absorbers along 16 GRB sightlines. The total
redshift path becomes $17.02$, and this results in a $dn/dz = 0.94$.
This surprising excess of strong {{Mg}{II}} absorbers with respect
to QSO sightlines remains a matter of debate and a satisfactory
explanation is still missing. Clearly more observations and analysis
are needed in order to solve this issue.


\end{document}